\documentclass[conference]{IEEEtran}
\IEEEoverridecommandlockouts
\usepackage{cite}
\usepackage{amsmath,amssymb,amsfonts}

\usepackage{graphicx}
\usepackage{textcomp}
\usepackage{xcolor}

\usepackage{times}
\usepackage{epsfig}
\usepackage{graphicx}
\usepackage{amsmath}
\usepackage{amssymb}
\usepackage{algorithm}
\usepackage{algpseudocode}

\usepackage{indentfirst}
\usepackage{multirow}
\usepackage{rotating}
\usepackage{array}
\usepackage{makecell}
\usepackage{tabularx}

\newcolumntype{C}[1]{>{\centering\arraybackslash}m{#1}}
\newcolumntype{Y}{>{\centering\arraybackslash}X}

\usepackage{bm}
\usepackage{amssymb}
\usepackage{pifont}

\usepackage{xcolor}
\usepackage{soul}
\usepackage{float}

\usepackage[caption=false, font=normalsize, labelfont=sf, textfont=sf]{subfig}

\def\BibTeX{{\rm B\kern-.05em{\sc i\kern-.025em b}\kern-.08em
    T\kern-.1667em\lower.7ex\hbox{E}\kern-.125emX}}

\begin{document}

\setlength{\textfloatsep}{4pt}
\setlength{\intextsep}{4pt}
\setlength{\floatsep}{4pt}

\title{Mapping Wireless Networks into Digital Reality through Joint Vertical and Horizontal Learning}

\author{
		\IEEEauthorblockN{Zifan Zhang$\IEEEauthorrefmark{1}$,
		Mingzhe Chen$\IEEEauthorrefmark{2}$,
            Zhaohui Yang$\IEEEauthorrefmark{3}$,
            Yuchen Liu$\IEEEauthorrefmark{1}$
 }
		\IEEEauthorblockA{
		$\IEEEauthorrefmark{1}$North Carolina State University, USA,
		$\IEEEauthorrefmark{2}$University of Miami, USA,
            $\IEEEauthorrefmark{3}$Zhejiang University, China
		}
	}

\maketitle
\begin{abstract}

In recent years, the complexity of 5G and beyond wireless networks has escalated, prompting a need for innovative frameworks to facilitate flexible management and efficient deployment. The concept of digital twins (DTs) has emerged as a solution to enable real-time monitoring, predictive configurations, and decision-making processes. While existing works primarily focus on leveraging DTs to optimize wireless networks, a detailed mapping methodology for creating virtual representations of network infrastructure and properties is still lacking. In this context, we introduce VH-Twin, a novel time-series data-driven framework that effectively maps wireless networks into digital reality. VH-Twin distinguishes itself through complementary vertical twinning (V-twinning) and horizontal twinning (H-twinning) stages, followed by a periodic clustering mechanism used to virtualize network regions based on their distinct geological and wireless characteristics. Specifically, V-twinning exploits distributed learning techniques to initialize a global twin model collaboratively from virtualized network clusters. H-twinning, on the other hand, is implemented with an asynchronous mapping scheme that dynamically updates twin models in response to network or environmental changes. Leveraging real-world wireless traffic data within a cellular wireless network, comprehensive experiments are conducted to verify that VH-Twin can effectively construct, deploy, and maintain network DTs. Parametric analysis also offers insights into how to strike a balance between twinning efficiency and model accuracy at scale.

\end{abstract}



\section{Introduction} \label{sec:intro}

In the realm of telecommunications, wireless networks are experiencing a paradigm shift, primarily driven by the advent of edge computing, spectrum sharing, and millimeter-wave communication technologies in 5G era. These technological advancements are foundational to a multitude of novel applications and services, notably enhancing mobile broadband and facilitating the seamless integration of Internet of Things (IoT)~\cite{nguyen20216g, qadir2023towards}, 
autonomous transportation~\cite{miglani2019deep}, urban infrastructure~\cite{cugurullo2020urban}, and remote healthcare delivery~\cite{qayyum2020secure}. Furthermore, the nascent stages of 6G research are indicative of potential revolutionary leaps in hybrid physical-virtual wireless technologies, paving the way for ubiquitous and intelligent connectivity worldwide.

Parallel to these advancements, the concept of digital twin (DT) has surfaced as a significant technological breakthrough~\cite{vanderhorn2021digital, semeraro2021digital, botin2022digital}. The DTs embody intricate virtual representations of physical entities or systems and gain traction in the context of the Fourth Industrial Revolution. This concept synergistically harnesses the capabilities of IoT, machine learning, and big data analytics, meticulously constructing a comprehensive digital model that mirrors the physical attributes, processes, and dynamics of its real-world counterpart. Such models play a pivotal role in facilitating predictive simulations, what-if analysis, and system optimizations within a virtual environment, thereby offering tangible insights into operational challenges and maintenance requirements~\cite{feng2023digital, yu2022energy,li2022digital}.

In essence, the foundation of DTs lies in a continuous mapping between the physical and digital domains. This convergence is exemplified by a process that intricately associates the physical properties of an entity or system with its digital realm. This concept, though evolving, has experienced significant advancements in recent study. For instance, \cite{tao2018digital, tao2022digital, boschert2016digital} explored the foundational aspects of DTs, establishing a comprehensive framework for their development and deployment across diverse sectors. These works highlight the importance of precise mapping for real-time synchronization, which is essential for predictive maintenance and operations of products. 
Furthermore, DT mappings extend beyond industrial applications to encompass urban planning and smart city initiatives, as demonstrated by~\cite{minerva2020digital} for high-stake ecosystems. In healthcare,~\cite{corral2020digital, jiang2020towards} emphasizing the versatility of DTs in personalizing patient care and advancing medical research. These studies collectively underscore the multifaceted nature of DT mappings as a cornerstone of emerging applications. In contrast, our work herein presents the inaugural attempt to adapt DT mapping techniques to wireless networks, aiming to comprehensively model the network system and attributes with heightened precision and efficacy.

In the telecommunication domains, existing works primarily focus on leveraging DTs to optimize wireless networks but often overlook the detailed mapping methodology for creating virtual representations of network infrastructure and properties~\cite{li2024JSTSP, khan2022digital, Yang2024ICASSP}. This mapping is a crucial prerequisite for subsequent applications of DTs in the field.
The construction of DTs in the context of wireless networks faces numerous challenges. A predominant obstacle lies in the integration of diverse data sources, crucial for accurately reflecting the complex dynamics of wireless networks. This integration involves processing a wide spectrum of sensory data, from network parameters to user behaviors and environmental factors, making it a data-intensive task. Furthermore, achieving real-time data synchronization between the physical network and its digital counterpart is imperative for maintaining an accurate model representation. This synchronization requires continuous processing and robust communication frameworks, posing both data exchange and resource-related challenges. Lastly, scalability also emerges as a significant concern during the mapping process. Given the mobility behavior of wireless devices, the DTs must be constantly updated and adapted to remain accurate, presenting a computation-demanding task due to the complexity and scale of contemporary network systems.

\begin{figure*}[ht]

    \centering
    \includegraphics[scale = 0.55]{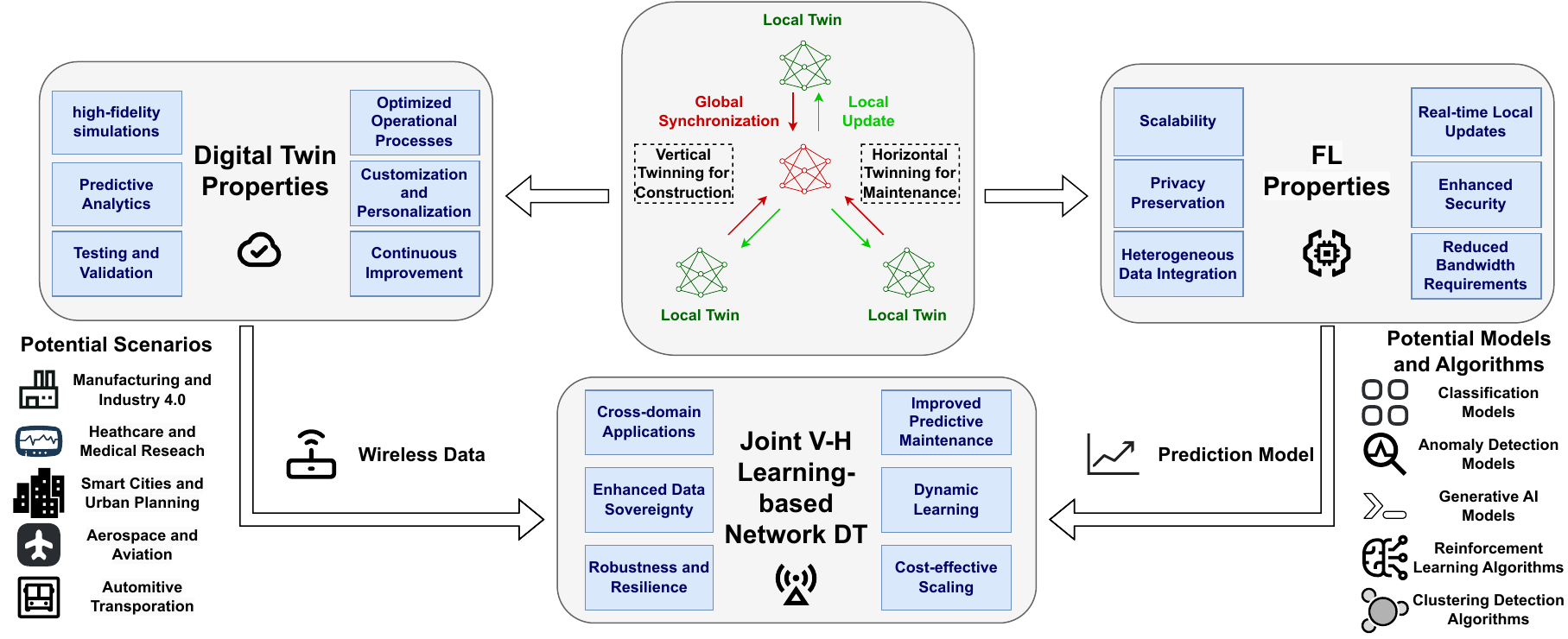}
    \caption{Motivation and basic architecture of network DT enabled by joint vertical and horizontal learning.}
    \label{fig:motivation}
    \vspace{-0.4cm}
\end{figure*}

To address the aforementioned challenges, our research pioneers a joint vertical-horizontal mapping methodology, called \textit{VH-Twin}, for the creation, deployment, and synchronization of DTs with time-series data in wireless networks. We first introduce a novel clustering algorithm, systematically executed to categorize network base stations (BSs) according to their distinct infrastructure characteristics. Following this, we exploit distributed machine learning techniques for precise digital twinning to ensure the privacy of handling time-series data within each BS, while maintaining overall mapping performance. As illustrated in Fig.~\ref{fig:motivation}, we explore the use of federated learning (FL), 
incorporating both synchronous and asynchronous approaches at different stages of the digital mapping process. Synchronous FL proves advantageous for its ability to facilitate consistent model updates and ensure uniform data distribution among participating BSs. In contrast, asynchronous FL offers increased flexibility in participation and timing of updates, catering to the dynamic nature of wireless environments. Accordingly, synchronous FL is employed to construct an initial twin model \textit{vertically}, while asynchronous FL is subsequently used to maintain the accuracy of the twin models \textit{horizontally} over time. The resulting self-evolved twins can then be applied for collaboratively forecasting precise time-series data, such as heterogeneous wireless traffic.
The main contributions are multifaceted and can be summarized as follows:
\begin{enumerate}
    \item We introduce a novel twinning framework that maps the wireless networks into digital reality, adaptively utilizing a time-series data-driven method for the creation and continuous synchronization of network DTs.
    
    \vspace{+0.1cm}
    \item We augment the twinning framework with a distinctive clustering algorithm that dynamically groups BSs based on both geographical and data communication characteristics. This enables the high-efficiency evolution of specialized twin models for each virtual network region.

    \vspace{+0.1cm}
    \item We demonstrate the feasibility of the proposed V-H Twin framework in addressing a real-world wireless traffic prediction problem within cellular wireless networks.

    \vspace{+0.1cm}
    \item Extensive evaluations show that VH-Twin can efficiently map the network attributes with heterogeneous wireless data, achieves adaptive twin synchronization, and scales effectively to accommodate dense wireless networks with parameterized configurations.
\end{enumerate}

\section{Preliminaries and Related Works} \label{sec:related}

\subsection{Digital Twin for Wireless Networks}

The integration of DTs in the wireless network sector represents a significant technological milestone within this rapidly evolving field. As detailed in~\cite{networkdigitaltwin,Y3090,irtf-nmrg-network-digital-twin-arch-05}, the application of DT involves constructing detailed virtual representations of network components and infrastructure. This facilitates real-time analytical capabilities and optimization processes, providing deeper insights into network behaviors across a variety of scenarios. Such methodologies are invaluable for predictive maintenance and performance monitoring, thereby substantially enhancing network reliability and efficiency.
In the realm of next-generation wireless technologies, the role of DT has become increasingly critical, as discussed in \cite{digitaltwin5g}. 
Through the simulation of diverse network configurations and load conditions, DT is investigated to 
provide the strategic deployment of 5G infrastructures, thereby maximizing coverage and data throughput.
\cite{connectingtwins} thoroughly examines the networking prerequisites for DT implementation. By emphasizing robust data collection systems and real-time data processing, it underscores the vital integration of IoT devices with DT to ensure a continuous and accurate flow of data for maintaining up-to-date virtual models. In \cite{graphneuralnetwork}, the innovative application of Graph Neural Networks in developing DT for network slicing is proposed, aiming at predicting network behaviors and dynamically optimizing resource management in high-demand bandwidth and low-latency scenarios.
Furthermore, the implementation of DT within vehicular networks is elucidated in \cite{intelligentdigitaltwin}. This research showcases the potential of DT in modeling and managing software-defined vehicular networks, significantly boosting the efficiency and reliability of vehicular communication systems. 
While prior works have explored versatile applications of DT in wireless networks, none of them explicitly delve into the creation and mapping process of network DTs, which is the primary focus of our work.

\subsection{Distributed Learning and DTs} 

In the rapidly evolving domain of DTs, distributed learning has emerged as a transformative approach, especially in the context of large-scale networks and Industrial IoT. The integration of these two technologies is not only enhancing system efficiency but also revolutionizing data processing capabilities. A prime example of this integration is seen in the adaptation of DTs for smart agricultural practices, as explored in~\cite{agri_digital_twin}. This study exemplifies the potential of DTs in real-time environmental monitoring and decision-making. 
\cite{adaptive_federated_iot} focuses on optimizing industrial IoT operations through the synergy of FL and DTs. This is further complemented by studies such as~\cite{cooperative_federated_blockchain}, which delve into the integration of blockchain technology for securing FL models within DT frameworks. 

Besides the data security enhancement, the challenge of efficient data communication in a distributed learning system is addressed in~\cite{communication_efficient_federated} and~\cite{communication_efficient_iot}. These studies propose to optimize data transfer and processing, a critical aspect of the scalability of massive access applications. 
Moreover, the dynamic nature of DTs, combined with incentive mechanisms in FL, is explored in~\cite{dynamic_air_ground}
for real-time data processing and accurate modeling. Lastly, the optimization of resource allocation using DT-enabled FL frameworks is investigated in~\cite{optimizing_federated_learning, resource_allocation_federated}, which demonstrates the effectiveness of using FL and DT in resource management problems. 
In contrast, our work explores the potential use of FL in addressing a fundamental DT mapping problem, laying the foundation for implementing a precise DT tailored for wireless networks. The distributed nature among multiple twins not only collaboratively handles the time-series data stream from diverse sources but also ensures real-time synchronization and scalability in dense wireless networks.

\subsection{Network Digital Twin for Traffic Forecasting}

As an application, we focus on a wireless traffic prediction problem as a use case to evaluate the efficacy of our mapped network digital twin (NDT) with time-series data streams. The architecture of this functional system is underpinned by an FL mechanism and encompasses a central DT, named global twin, coordinating with a network of \( M \) BSs. Each BS, denoted as \( m \) in the set \([M]\), independently holds a proprietary dataset \( d_m = \{d^1_m, d^2_m, \ldots, d^L_m\} \). In this dataset, \( L \) indicates the total number of time intervals, and \( d^l_m \) represents the traffic load at BS \( m \) during the \( l \)-th interval, where \( l \) ranges over \([L]\).

The NDT involves constructing input-output predictive traffic sequences locally, denoted as \( \{r_m^n, s_m^n\}_{n=1}^{z} \), for each BS to generate future predictions. Here, \( r_m^n \) is a subset of historical traffic data corresponding to the output \( s_m^n =\{d_m^{l-1}, \dots, d_m^{l-a}, d_m^{l-\rho 1},\dots, d_m^{l-\rho b}\} \). The parameters \( a \) and \( b \) represent sliding windows that capture immediate and cyclical temporal dependencies, respectively, while \( \rho \) reflects inherent periodicities in the network, which might be influenced by user activity patterns or application service demands.

With the real-time synchronization with the physical networks, the NDTs can be tailored for one-step-ahead traffic forecasts. Specifically, for each BS \( m \), their corresponding NDT predicts the upcoming traffic load \( \tilde{s}^n_m \) using the historical data \( r^n_m \) and a twin model vector \( \bm{\alpha} \). This prediction is formulated as \( \tilde{s}^n_m = f(r^n_m, \bm{\alpha}) \), where \( f(\cdot) \) represents the regression function.

Considering an FL-based NDT system for traffic forecasting, the primary objective is to minimize prediction errors across all \( M \) BSs for a better understanding of the physical network. This can be formulated as an optimization problem, aiming to identify the optimal DT model \( \bm{\alpha}^* \) for the downstream tasks. The problem is generally defined as:
\[
\bm{\alpha}^* = \arg\min_{\bm{\alpha}}  \frac{1}{Mz} \sum_{m=1}^{M} \sum_{n=1}^{z} F(f(r_m^n, \bm{\alpha}), s_m^n),
\]
with \( F \) being the quadratic loss function, specifically \( F(f(r_m^n, \bm{\alpha}), s_m^n) = \left| f(r_m^n, \bm{\alpha}) - s_m^n \right|^2 \). This problem can be resolved through FL with distributed NDTs, and unfolds in each global training round \( t \) as follows:

\begin{enumerate}
    \item \textit{Synchronization:} The global twin broadcasts the current twin model \( \bm{\alpha}^t \) to all participating BSs.
    \item \textit{Local Updating:} Each BS \( m \), within the range \( M \), refines its local twin using its private dataset and the current global twin model. The updated local twin, \( \bm{\alpha}_m^{t} \), is then sent back to the global twin.
    \item \textit{Models Aggregation:} The global twin aggregates the attributes from the \( M \) connected local twins using an aggregation rule (AGG), and updates its own functionality as \( \bm{\alpha}^{t+1} = \text{AGG}\{\bm{\alpha}_1^{t}, \bm{\alpha}_2^{t}, \ldots, \bm{\alpha}_M^{t}\} \). Typically, Federated Averaging (FedAvg) method~\cite{mcmahan2017communication} can be adopted in this problem, where the global twin directly average the model parameters of \( M \) local twins for evolution, expressed as \(\text{AGG}\{\bm{\alpha}_1^{t}, \bm{\alpha}_2^{t}, \ldots, \bm{\alpha}_M^{t}\} = \frac{1}{M}\sum_{m=1}^M \bm{\alpha}_m^{t}\).
\end{enumerate}

This single-level mapping approach utilizes a classical FL strategy to aggregate multiple local twin models, and we will employ this scheme as the baseline in Sec. IV to compare it with our proposed joint vertical-horizontal mapping scheme.

\section{Joint Vertical and Horizontal Twinning for Digital Network Mapping} 
\label{sec:system_design}

To achieve a high-fidelity and high efficient digital network mapping, we introduce an innovative framework termed VH-Twin, incorporating both vertical twinning, which involves initial mapping from physical networks to create a global network digital twin (G-NDT), and horizontal twinning, which integrates with legacy digital twins for synchronization.
This framework is also wrapped with a dynamic connectivity segmentation (DCS) stage, which is employed periodically to ensure effective regional-based network digitalization. 

\begin{figure}[h]
    \centering
    \vspace{-0.0cm}
    \includegraphics[scale = 0.34]{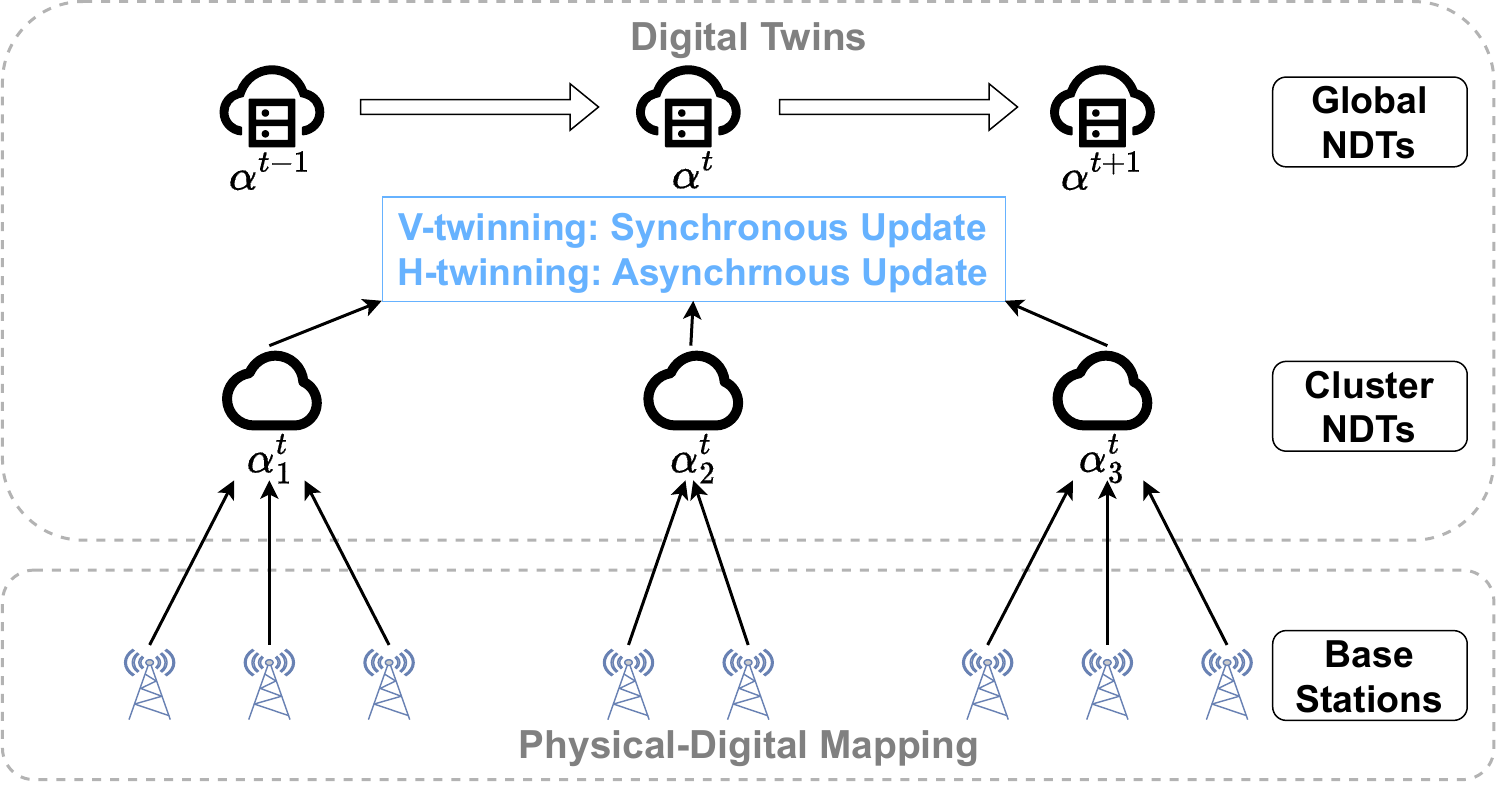}
    \vspace{-0.1cm}
    \caption{Framework of VH-Twin.}
    \label{fig:framework}
    \vspace{-0.0cm}
\end{figure}

\begin{figure}[h]
    \centering
    \vspace{-0.0cm}
    \includegraphics[scale = 0.65]{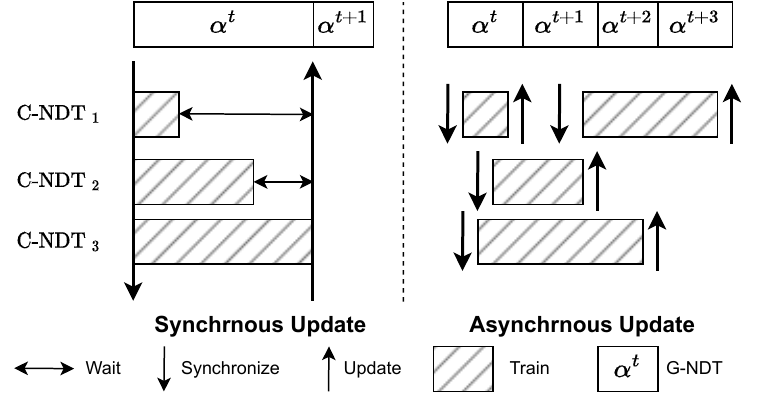}
    \vspace{-0.1cm}
    \caption{Synchronous and asynchronous NDT mappings.}
    \label{fig:FL}
    \vspace{-0.0cm}
\end{figure}

\subsection{Dynamic Connectivity Segmentation}

\begin{algorithm}[h]
\caption{Dynamic Connectivity Segmentation (DCS)}
\begin{algorithmic}[1]
\Require Relationship matrix \( \{ g, k, \beta, \tau \} \), attribute weights \( \omega \), \( C \) clusters, \( N \) BSs
\Ensure Clusters \( c \)
\State Initialize \( \Phi \);
\For{\( m_1 = 1 \) to \( N \)}
    \For{\( m_2 = 1 \) to \( N \)}
        \State \( \Phi_{m_1,m_2} \gets \omega_g / g_{m_1,m_2} + \omega_k \cdot k_{m_1,m_2} \)
        \State \( ~+ \omega_\beta \cdot \beta_{m_1,m_2} + \omega_\tau \cdot \tau_{m_1,m_2} \)
    \EndFor
\EndFor
\If{Fixed \( C \) clusters desired}
    \State Identify clusters by \( \Phi \) with Girvan-Newton method
\ElsIf{Adaptive \( C \) clusters desired}
    \State Identify clusters by \( \Phi \) with Louvain method
\EndIf

\State \Return \( c \)
\end{algorithmic}
\label{algo:DCS}
\end{algorithm}

The DCS algorithm, as outlined in Algorithm~\ref{algo:DCS}, is designed to cluster multiple BSs responsible for network service areas with similar communication characteristics and networking configurations. This clustering step is integral to the following creation and updates of multiple distributed network DTs, i.e. clustered network digital twins (C-NDT), which demonstrate distinct network behaviors and perform paralleled synchronization with the G-NDT.
This algorithm is executed periodically, ensuring dynamic clustering and thereby enhancing the twinning performance in real time. 
In the DCS algorithm, clusters are formed based on an attribute sequence of BSs \( \{ g, k, \beta, \tau \} \), representing their geological distances, backhaul link capacities to the core network, coverage area overlaps, and similarity of wireless traffic data distribution, respectively. 
Initially, a relationship matrix incorporating these attributes, weighted by \( \omega \), is constructed. For instance, the algorithm calculates a metric \( \Phi_{m_1,m_2} \) to quantify the correlation between BS \( m_1 \) and BS \( m_2 \) using the formula as:
\begin{equation}
    \Phi_{m_1,m_2} = \frac{\omega_g}{g_{m_1,m_2}} + \omega_k \cdot k_{m_1,m_2} + \omega_\beta \cdot \beta_{m_1,m_2} + \omega_\tau \cdot \tau_{m_1,m_2},
\end{equation}
where $\omega$ is the tunable weight to balance the significance of each network attribute. 
Next, to assess interconnection strength among BSs within a wireless network, two clustering methods are employed based on different scenarios. The first, utilized when a specific cluster count \( C \) is needed, involves calculating \textit{betweenness centrality} denoted by \( \Phi \). This measure indicates how frequently an edge serves as a bridge on the shortest network paths. Following a Girvan-Newman-like approach~\cite{Newman_2004}, we iteratively remove edges with low \( \Phi \) until the desired cluster count \( C \) is achieved.
For scenarios requiring flexible cluster numbers, we employ the Louvain-like method~\cite{Blondel_2008}, which maximizes network modularity by initially assigning each node to an individual community and then iteratively merging these communities to enhance the overall modularity score. The method assesses the edge density within communities relative to what is expected in a random graph. Specifically, a high modularity score indicates a strong community division, characterized by dense internal connections within communities and sparser links between them. In essence, this approach allows for the identification of the network's hierarchical community structure, unveiling both minor and major community clusters. On the other hand, there is no need to specify the number of clusters, making it a more adaptive approach for the subsequent twinning process.
Overall, this clustering process is crucial in effectively virtualizing network regions covered by distributed BSs into several C-NDTs,
ultimately contributing to the update of G-NDT, as discussed in Sec. III-B-C.

\subsection{Vertical Twinning for Initialization}

\begin{algorithm}[h]
\caption{Vertical Twinning (V-twinning)}
\begin{algorithmic}[1]
\Require Local twin models \( \bm{\alpha}_1^t, \bm{\alpha}_2^t, \ldots, \bm{\alpha}_{M}^t \), \( C \) clusters
\Ensure Updated G-NDT \( \bm{\alpha}^{t+1} \)
\For{\( c = 1, 2, \ldots, C \textbf{ synchronously}\)}
    \State \( P \gets \text{BS in cluster } c \)
    \State \(\bm{\alpha}_c^{t} \gets \frac{1}{P}\sum_{p=1}^P \bm{\alpha}_p^{t} \)
\EndFor
\State \( \bm{\alpha}^{t+1} \gets \frac{1}{C}\sum_{c=1}^C \bm{\alpha}_c^{t}\)
\State \Return \( \bm{\alpha}^{t+1} \)
\end{algorithmic}
\label{algo:VT}
\end{algorithm}

The V-twinning stage aims to create an initial G-NDT with historical time-series data on, e.g., wireless traffic from the physical network. It employs an FL-based framework for accurate mappings, specifically tailored for wireless networks with multiple BSs organized in clusters. FL is well-suited for this problem because model parameters are shared among BSs instead of raw data, enabling collaborative training of a global model. This approach efficiently distributes twinning tasks across BSs while ensuring content data privacy. As depicted in Algorithm~\ref{algo:VT}, historical time-series data from each BS \( m \) are used to train C-NDTs for each cluster \( c \). With local twin models shared from BSs within the same cluster, denoted as \(\bm{\alpha}_1^t, \bm{\alpha}_2^t, \ldots, \bm{\alpha}_M^t\), the corresponding C-NDT \(\bm{\alpha}_c^t\) aggregates the models to reach a consensus. The most common aggregation rule FedAvg~\cite{mcmahan2017communication} can be used to compute the dimension-wise arithmetic mean of each twin model parameter.

G-NDT, represented as \(\bm{\alpha}^{t+1}\), is the averaged aggregator of multiple C-NDTs at $C$ clusters, i.e.,  \(\bm{\alpha}^{t+1} = \frac{1}{C}\sum_{c=1}^C \bm{\alpha}_c^t\), where \(C\) is the number of clusters. Then, the model parameters of G-NDT are sent back to each cluster for synchronizing C-NDTs after the twinning aggregation process.

Specifically, the V-twinning process employs \textit{synchronous} NDT mapping update, ensuring that all C-NDTs update their twin models simultaneously. 
It involves a coordinated training process where all participating twins update their local models and synchronize these updates with a central server, typically at predetermined intervals, as shown in Fig.~\ref{fig:FL}.
This synchronicity is crucial during the initial twinning process, which always demands a large amount of data from the physical counterpart to build a consistent twin basis across the network regions. 
Additionally, the use of synchronous mapping updates can simplify the management of model updates and reduce issues related to stale or incompatible data, making it suitable for network scenarios where uniformity and coordination among BSs are critical.

\subsection{Horizontal Twinning for Model Evolution}

\begin{algorithm}[h]
\caption{Horizontal Twinning (H-twinning)}
\begin{algorithmic}[1]
\Require Local twin models \( \bm{\alpha}_1^t, \bm{\alpha}_2^t, \ldots, \bm{\alpha}_{M}^t \), current G-NDT \( \bm{\alpha}^{t} \), \( C \) clusters, threshold \( \psi \)
\Ensure Updated G-NDT \( \bm{\alpha}^{t+1} \)
\For{\( c = 1, 2, \ldots, C \textbf{ asynchronously}\)}
    \State \( P \gets \text{BS in cluster } c \)
    \State \(\bm{\alpha}_c^{t} \gets \frac{1}{P}\sum_{p=1}^P \bm{\alpha}_p^{t} \)
    \State \( \epsilon \gets (\bm{\alpha}_c^{t} + \bm{\alpha}^{t})^2 \)
    \If{\( \epsilon > \psi \)}
        \State \( \bm{\alpha}^{t+1} \gets \frac{1}{C}\sum_{c=1}^C \bm{\alpha}_c^{t}\)
    \Else
        \State \( \bm{\alpha}^{t+1} \gets \bm{\alpha}^{t}\)
    \EndIf
\EndFor
\State \Return \( \bm{\alpha}^{t+1} \)
\end{algorithmic}
\label{algo:HT}
\end{algorithm}

To ensure the network DTs remain relevant in a dynamic wireless environment, H-twinning stage is designed to periodically synchronize between the physical network and DTs with time-series data. Unlike V-twining, it adopts an \textit{asynchronous} NDT mapping strategy to update with fluctuations from the physical network, aiming to provide a scalable and flexible solution for wireless networks composed of multiple clusters.

As described in Algorithm \ref{algo:HT}, H-twinning begins with $N$ local models \(\bm{\alpha}_1^t, \bm{\alpha}_2^t, \ldots, \bm{\alpha}_M^t\) from respective BSs and the current G-NDT \(\bm{\alpha}^t\). The threshold \(\psi\) serves as a criterion to decide whether the G-NDT should be updated. It assesses the deviation between a C-NDT \(\bm{\alpha}_c^t\) and the current G-NDT \(\bm{\alpha}^t\), quantified by \(\epsilon = (\bm{\alpha}_c^t - \bm{\alpha}^t)^2\). If \(\epsilon\) surpasses the threshold \(\psi\), indicating a significant change in the physical network, the G-NDT is updated to reflect the fresh information. The updated G-NDT, \(\bm{\alpha}^{t+1}\), is calculated as an average of C-NDTs at the current time slot, \(\bm{\alpha}^{t+1} = \frac{1}{C}\sum_{c=1}^C \bm{\alpha}_c^t\). If \(\epsilon\) is within the threshold \(\psi\), the G-NDT remains with the current model, i.e. \(\bm{\alpha}^{t+1} = \bm{\alpha}^t\). This threshold-based update mechanism enhances the network's efficiency by ensuring that only significant changes will lead to twin updates, thereby reducing unnecessary computational overhead and preserving bandwidth.

Compared with V-twinning, H-twinning does not require simultaneous updates from all clusters, although the overall procedure to compute C-NDT and G-NDT is quite similar. 
In the context of this hybrid time-series learning, the model update mechanisms differ significantly between synchronous updates and asynchronous updates. In synchronous updates, the global model update is synchronized across all participating twins at specific intervals. Thus, the G-NDT at time \( t \) can be updated as follow:
\begin{equation}
    \bm{\alpha}^{t} = \frac{1}{C} \sum_{c=1}^{C} \bm{\alpha}_c^{t}.
\end{equation}
Here, \( \bm{\alpha}^{t} \) represents the G-NDT model after the \( t \)-th synchronization, \( C \) is the total number of clusters, and \( \bm{\alpha}_c^{t} \) are the C-NDT model parameters from cluster \( c \) at the \( t \)-th synchronization. 
On the other hand, asynchronous updates enable each twin to update the global twin model based on its individual training schedule. When each C-NDT independently contributes its twinning results that cover specific network regions, the G-NDT can be updated continuously as:
\begin{equation}
    \bm{\alpha}^{t_n} = \bm{\alpha}^{t_{n-1}} + \eta_n (\bm{\alpha}_c^{t_n} - \bm{\alpha}^{t_{n-1}}).
\end{equation}
In this equation, \( \bm{\alpha}^{t_n} \) is the G-NDT model updated after the \( c \)-th twin's contribution at its local time \( t_n \), \( \eta_c \) is a weighting factor for the \( c \)-th twin, \( \bm{\alpha}_c^{t_n} \) are the C-NDT model parameters from twin \( c \) at its local time \( t_n \), and \( \bm{\alpha}^{t_{n-1}} \) is the G-NDT model before the \( n \)-th update. This approach enables continuous and potentially faster adaptation but requires careful management of update consistency.
It allows each participating twin to update and share its model updates with the central server independently and at different times, without waiting for other C-NDTs, thereby accommodating diverse data availability and computational capacities across the network.
In a practical scenario, each BS \( m \) can store the real-time data stream and then train its affiliated C-NDT model when the amount of data reaches a threshold in batches. The asynchronous mapping mechanism allows for greater flexibility in participation, as BSs and C-NDTs within edge servers may contribute to the twinning process at their own pace and availability, without being bound to a strict synchronization schedule. This feature is particularly beneficial in wireless networks with BSs having varying computational resources or backhaul connectivity, ensuring that the twinning process is inclusive and efficient even in less ideal network conditions.


\section{Evaluation Results and Analysis} \label{sec:exp}

In this section, we perform comprehensive evaluations to assess the performance of VH-Twin in a wireless traffic forecasting scenario, aiming at constructing, deploying, and maintaining precise twin models. The dataset and code implementation is available at~\cite{dnt}.

\subsection{Dataset}
 
In this evaluation, we employ a real-world dataset provided by Telecom Italia, as referenced in~\cite{dataset}, to evaluate the effectiveness of our proposed VH-twin. The dataset, originating from Milan City, is composed of wireless traffic data intricately divided into 10,000 grid cells. Each cell, measuring approximately 235 meters on each side, is equipped with its own dedicated BS. These BSs collectively capture a wealth of time-series data records, which include detailed information on short message service (SMS), call activities, and internet usage patterns. This extensive dataset facilitates a comprehensive analysis of urban telecommunications behaviors. Specifically, our focus centers on utilizing internet usage data to construct traffic twin models with the VH-Twin framework.

\subsection{Performance Metrics}

To evaluate the precision of the constructed twins, we use Mean Squared Error (MSE), Mean Absolute Error (MAE), and Normalized Root Mean Square Error (NRMSE) as key performance metrics. These statistical measures indicate distinctions between the time-series traffic data collected from the physical cellular network and that generated from the mapped NDTs. offering insights into the model discrepancy between the twins and their real-world counterparts. Lower values of MSE, MAE, and NRMSE indicate higher precision and, consequently, more effective performance of our V-twinning and H-twinning processes.
Additionally, we evaluate the efficiency of the V-twinning process by measuring the initial model training time 
within a FL framework. This metric enables a comparison of the time duration required for model initialization during the V-twinning stage against that of a baseline \textit{single-level mapping scheme}. In the single-level mapping scheme, local twin models are constructed on each BS without clustering and aggregated using the same aggregation rule, as detailed in Sec. II-C, to generate a global twin model, and the synchronous updating is enforced in vertical and horizontal mapping stages.

Next, to evaluate the overhead of twin evolution during the subsequent model update process, we analyze the communication round required for the H-twinning phase, indicating the cost of data exchanges needed to update G-NDT when necessary.
This measurement is crucial for understanding the network bandwidth and time costs associated with maintaining the fidelity of NDTs in a dynamic wireless environment. Particularly, both model training time and communication rounds are normalized by dividing by 10,000 for graphical representation in this paper.

\subsection{Network and Experiment Settings}
In our evaluation, we randomly select 50 BSs in an urban network environment to assess the performance of our joint DCS, V-twinning, and H-twinning methodologies. In the DCS phase, we implement the clustering algorithm at different intervals for the vertical and horizontal twinning stages, respectively -- every 10 model training epochs out of a total of 100 for V-twinning, and every 5 model training epochs out of 20 for H-twinning. 
We set up the default network topology with distributed BSs as a 20-regular graph, implying that each BS is connected to 20 neighbors. Although the connections maintain a consistent number, the backhaul link strengths and other attributes may vary over time. We ensure a balanced distribution of weight across each attribute to maintain equilibrium.
Lastly, for FL-based model training stage, we set the learning rate at 0.01 and choose a batch size of 64. The FedAvg rule serves as the default model aggregation method.

\subsection{Numerical Evaluations}

\begin{figure}[h]
    \centering
    \vspace{-0.2cm}
    \includegraphics[scale = 0.25]{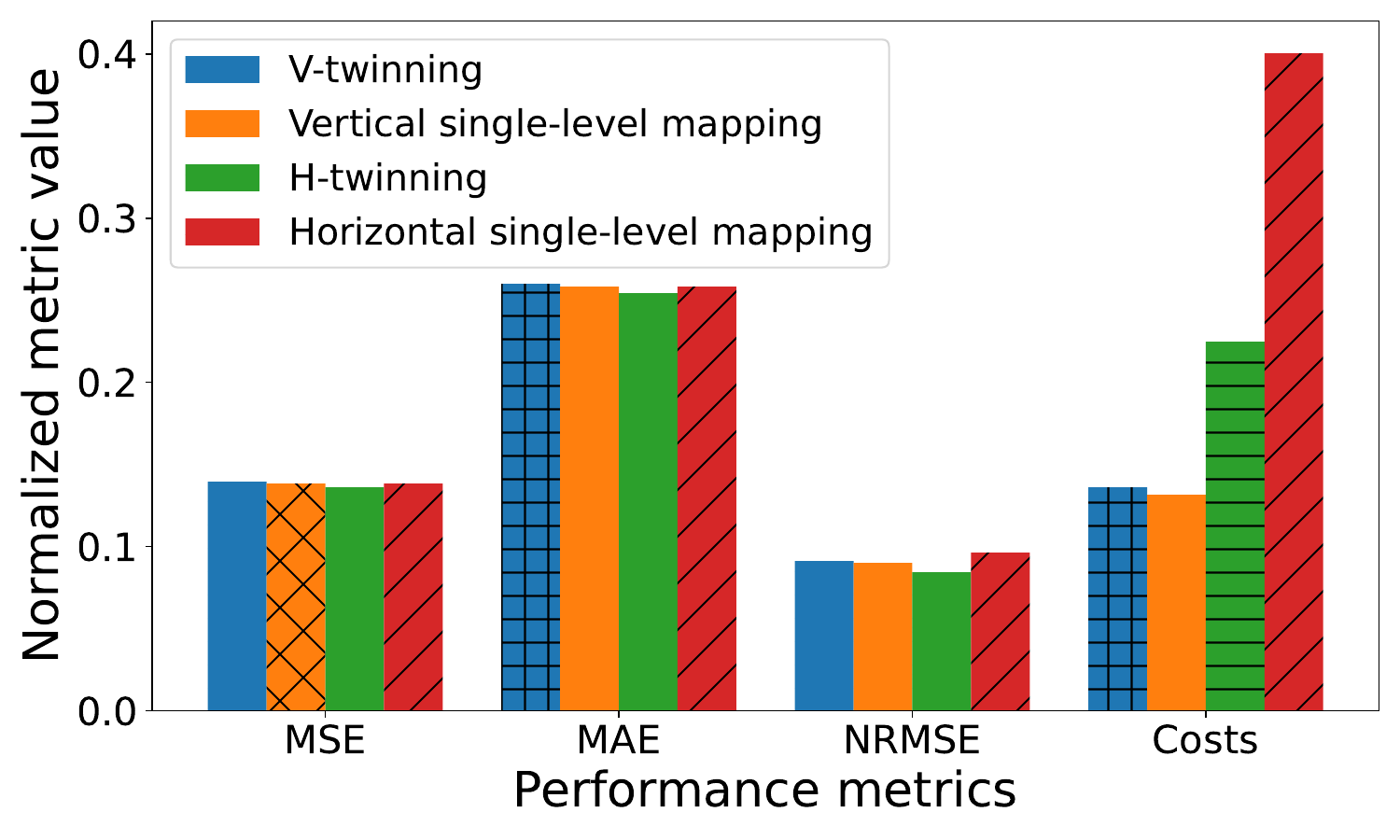}
    \vspace{-0.4cm}
    \caption{Performance metrics of VH-Twin compared with single-level twinning.}
    \label{fig:vh}
    \vspace{-0.1cm}
\end{figure}

\subsubsection{\textbf{Accuracy and efficiency of VH-Twin}}

We evaluate the accuracy of constructed twin models and mapping efficiency using our VH-Twin. The single-level twinning method is also included as a point of comparison, where the results are reported in Fig.~\ref{fig:vh}. 
For the single-level mapping, we use classical FL to aggregate local twin models without clustering. 
In the vertical twinning stage, MSE, MAE, and NRMSE for single-level twinning are observed to be 0.138, 0.258, and 0.090, respectively. By contrast, the error rates of our VH-Twin at this stage are marginally elevated, exceeding the single-level twinning by approximately 1.5\%. This occurs because models developed via clustering tend to excel within their designated clusters but often face challenges in effectively generalizing to different groups or the entire dataset.
Additionally, the twinning overhead for VH-Twin exhibits a modest increase over the single-level twinning, estimated at 3.5\%. This increment can be attributed to the additional DCS stage in our VH-Twin, which first virtualizes network regions covered by BSs into distinct clusters. Following that, NDTs are tasked with aggregating local twin models from the BSs and dispatching them back, thereby potentially extending the overall twinning process. 
However, theoretically, as VH-Twin exclusively utilizes synchronous updates, the communication rounds should be comparatively analogous to those of single-level twinning, which aligns with the evaluation results obtained herein.

Despite the slight compromise in the one-time V-twinning stage, the full potential of VH-Twin is revealed during the subsequent H-twinning stage. First, our H-twinning process showcases an enhancement in twin accuracy when compared to single-level twinning method. This is because asynchronous model updates can effectively handle real-time heterogeneous data, resulting in a more stable and accurate twin as the time-series data evolves.
Notably, the required overhead for H-twinning is significantly reduced in comparison to single-level twinning, by approximately 43.8\%. Due to the slightly prolonged initialization of twin models and the significantly reduced synchronization costs with the physical network, our joint vertical-horizontal twinning scheme yields a more efficient framework for digital network mapping. 

\begin{table*}[!ht]
\centering
\caption{Impact of Cluster Configurations on V-twinning and H-twinning Processes}
\begin{tabular}{|c|c|c|c|c|c|c|c|c|}
\hline
\multirow{2}{*}{\textbf{Metric}} & \multicolumn{2}{c|}{\textbf{5 Clusters}} & \multicolumn{2}{c|}{\textbf{10 Clusters}} & \multicolumn{2}{c|}{\textbf{20 Clusters}} & \multicolumn{2}{c|}{\textbf{30 Clusters}} \\ \cline{2-9} 
                                 & V-twinning           & H-twinning           & V-twinning            & H-twinning            & V-twinning            & H-twinning            & V-twinning            & H-twinning            \\ \hline
MSE                              & 0.140                & 0.140                & 0.139                 & 0.136                 & 0.140                 & 0.035                 & 0.141                 & 0.135                 \\ \hline
MAE                              & 0.260                & 0.258                & 0.260                 & 0.256                 & 0.260                 & 0.253                 & 0.262                 & 0.256                 \\ \hline
NRMSE                            & 0.088                & 0.089                & 0.091                 & 0.089                 & 0.092                 & 0.088                 & 0.091                 & 0.088                 \\ \hline
Initial Mapping (sec.)                             & 1356                & N/A                & 1359                 & N/A                & 1368                 & N/A                & 1372                 & N/A                \\ \hline
Update (rounds)                             & N/A                & 2169                & N/A                & 2360                 & N/A                & 2921                 & N/A                & 3164                 \\ \hline
\end{tabular}
\label{table:cluster}
\vspace{-0.3cm}
\end{table*}

\vspace{+0.05cm}
\subsubsection{\textbf{Impact of Cluster Configuration on VH-Twin}}

As explained in Sec. III, the clustering process is critical in effectively virtualizing network regions covered by distributed BSs, providing a solid foundation for the subsequent twinning process. Utilizing the Girvan-Newton clustering algorithm to appropriately group BSs into the desired number of clusters, Table~\ref{table:cluster} shows the variations in twinning accuracy and efficiency across different cluster configurations. 

First, it is observed that the accuracy of twin models, encapsulated by MSE, MAE, and NRMSE, exhibits minimal fluctuations across different cluster sizes for the V-twinning stage. Specifically, MSE and MAE values remain relatively stable across 5, 10, 20, and 30 clusters, thereby maintaining a similar performance trend. This aligns with the common sense that global model errors should not be significantly impacted by the number of clusters when collecting the initial data from BSs. Nevertheless, it is observed that lower cluster counts in the H-twinning stage result in reduced error rates, potentially attributable to the diminished data heterogeneity at each BS.
When it comes to the efficiency section, there is a noticeable rise in initial mapping time for V-twinning as the number of clusters increases. This trend is evident from the progression of model training time, which escalates from 1,356 seconds for 5 clusters to 1,372 seconds for 30 clusters. Such an increase can be attributed to the enhanced computational load and complexity associated with managing more clusters.
Regarding the cost for twin update process, measured by communication rounds between C-NDTs and G-NDT, the H-twinning process results in a progressive increase in communication overhead as the number of clusters rises. This is evidenced by the twin model update of 2,169 seconds for 5 clusters, escalating to 3,164 seconds for 30 clusters. This suggests that more clusters lead to higher twin synchronization overhead, which could potentially result in an inclination towards local model training due to efficiency concerns.
It is also noteworthy that the mapping time metric is not applicable (N/A) for H-twinning, and the update round is not applicable for V-twinning. This distinction highlights the different operational purposes of the two twinning phases and underscores the trade-off between data heterogeneity and overall twinning overhead when choosing an appropriate cluster configuration.

\begin{table*}[!ht]
\centering
\caption{Impact of BS Density on V-twinning and H-twinning Processes}
\begin{tabular}{|c|c|c|c|c|c|c|c|c|}
\hline
\multirow{2}{*}{\textbf{Metric}} & \multicolumn{2}{c|}{\textbf{25 BSs}} & \multicolumn{2}{c|}{\textbf{50 BSs}} & \multicolumn{2}{c|}{\textbf{100 BSs}} & \multicolumn{2}{c|}{\textbf{200 BSs}} \\ \cline{2-9} 
                                 & V-twinning           & H-twinning           & V-twinning            & H-twinning            & V-twinning            & H-twinning            & V-twinning            & H-twinning            \\ \hline
MSE                              & 0.143                & 0.120                & 0.139                 & 0.134                 & 0.139                 & 0.138                 & 0.135                 & 0.135                 \\ \hline
MAE                              & 0.270                & 0.248                & 0.260                 & 0.254                 & 0.260                 & 0.259                 & 0.261                 & 0.261                 \\ \hline
NRMSE                            & 0.092                & 0.085                & 0.091                 & 0.087                 & 0.091                 & 0.091                 & 0.090                 & 0.090                 \\ \hline
Initial Mapping (sec.)                             & 642                & N/A               & 1359                 & N/A                & 2628                 & N/A                & 6665                 & N/A                \\ \hline
Update (rounds)                             & N/A               & 3000                & N/A                & 2260                 & N/A                & 2325                 & N/A                & 2302                 \\ \hline
Number of clusters                         & 4               & 4                & 5                & 5                 & 6                & 6                 & 10                & 10                 
\\ \hline
\end{tabular}
\label{table:count}
\vspace{-0.3cm}
\end{table*}

\vspace{+0.05cm}
\subsubsection{\textbf{Impact of BS density on mapping process}}

To evaluate the scalability of VH-Twin in accommodating dense wireless networks, Table~\ref{table:count} presents a comprehensive overview of how the deployment density of BSs affects our twinning process. The analysis reveals that increasing the number of BSs from 25 to 200 in a large urban environment has a negligible impact on twinning accuracy. This is evident from the stable MSE, MAE, and NRMSE across different BS densities. Both V-twinning and H-twinning stages exhibit minor fluctuations, maintaining a consistent performance trend. These findings suggest the robustness of VH-Twin to changes in network scale, ensuring reliable model accuracy.
As expected, there is a notable increase in the mapping time for V-twinning as the network scale expands. The mapping time rises from 642 seconds with 25 BSs to 6,665 seconds with 200 BSs, reflecting the necessity to gather more data for the initial twinning process. 
In contrast, 
the update rounds during the H-twinning period show a reduction with an increased BS density, albeit with a less steep curve, ranging from 3,000 seconds for 25 BSs to 2,302 seconds for 200 BSs. This decrease in mapping overhead suggests that a greater number of BSs could lead to more decent initial model accuracy and stability. Consequently, the G-NDT may not require frequent updates to synchronize with the physical networks.
Furthermore, an insightful finding uncovered through the data analysis presented in Table~\ref{table:count} pertains to our adaptive clustering mechanism based on the Louvain method, where the number of clusters expands alongside the increase in network scale for both V-twinning and H-twinning processes. This outcome underscores the potential of the adaptive clustering stage to alleviate complexity and reduce twin update overhead (i.e., in the H-Twinning phase), particularly in the context of large-scale networks. Such pre-processing contributes to ensuring the scalability and efficiency of the mapping process.

\begin{figure}[h]
    \vspace{-0.0cm}
    \centering
    \includegraphics[scale = 0.25]{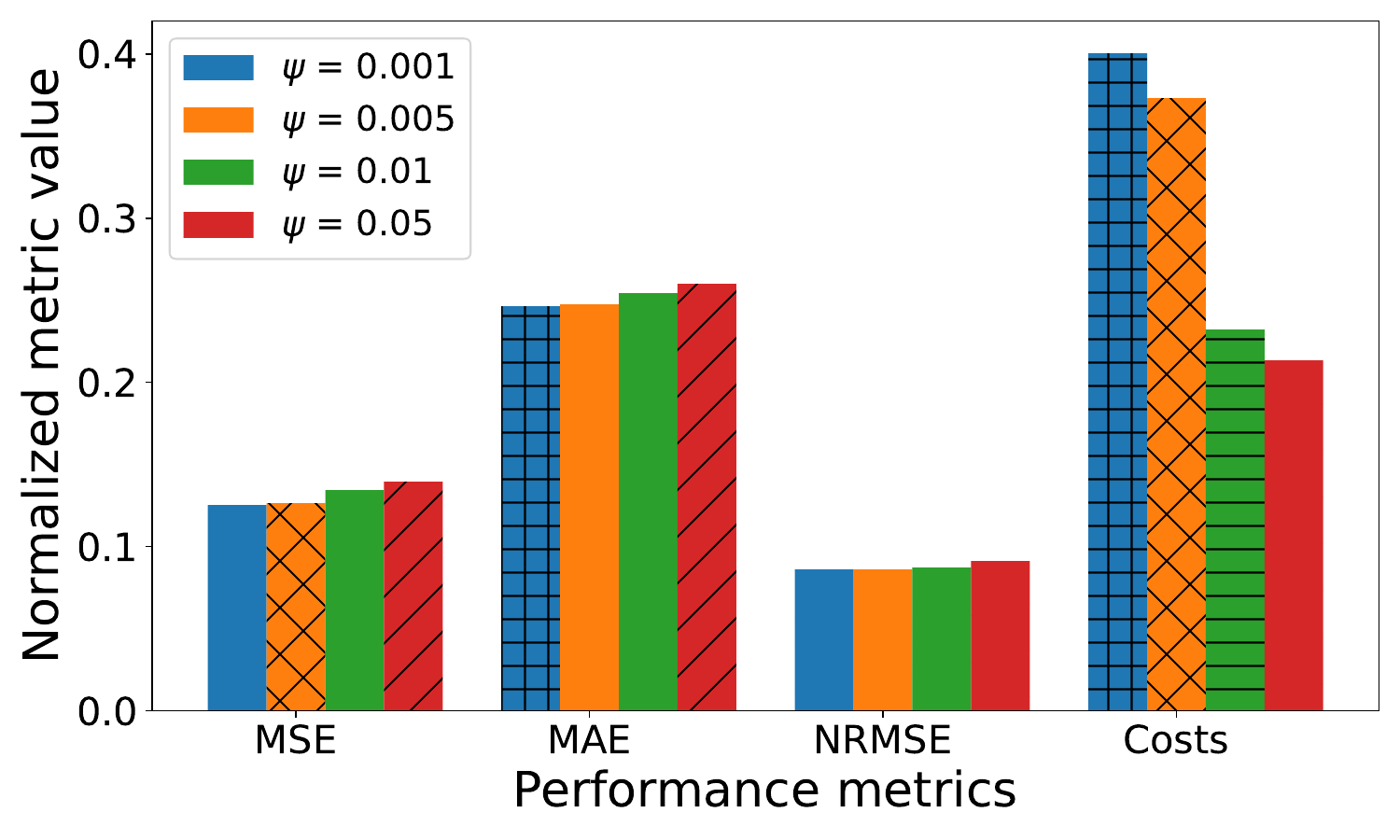}
    \vspace{-0.3cm}
    \caption{Impact of different values of \( \psi \).}
    \label{fig:threshold}
    \vspace{-0.0cm}
\end{figure}

\subsubsection{\textbf{Trade-off between mapping accuracy and efficiency}}
In Algorithm 3 for the H-Twinning stage, a tunable threshold value \( \psi \) is utilized to regulate the frequency of twin model updates. If the disparity between a current C-NDT and the G-NDT exceeds \( \psi \), the G-NDT will synchronize with the corresponding C-NDT for model evolution. Otherwise, the G-NDT remains unchanged. 
Fig.~\ref{fig:threshold} presents a comprehensive insight into how different values of \( \psi \) influence the performance of twin model. As \( \psi \) increases from 0.001 to 0.05, a notable trend of rising model error is observed, suggesting a decrease in twinning accuracy with higher threshold values due to the reduced synchronization over time.
As expected, a significant observation from Fig.~\ref{fig:threshold} is the reduction in mapping costs with larger thresholds, primarily attributed to the decrease from 4,000 at the lowest threshold to 2,132 at the highest. These results highlight a critical trade-off between mapping efficiency and model accuracy achieved through the tuning of \( \psi \). For instance, there is potential to improve twinning overhead by up to 50\% by sacrificing approximately 10\% accuracy in our evaluated scenario.

\begin{table*}[!ht]
\centering
\caption{Impact of Percentage of Mapped BSs on V-twinning and H-twinning Processes}
\begin{tabular}{|c|c|c|c|c|c|c|c|c|}
\hline
\multirow{2}{*}{\textbf{Metric}} & \multicolumn{2}{c|}{\textbf{40\%}} & \multicolumn{2}{c|}{\textbf{60\%}} & \multicolumn{2}{c|}{\textbf{80\%}} & \multicolumn{2}{c|}{\textbf{100\%}} \\ \cline{2-9} 
                                 & V-twinning           & H-twinning           & V-twinning            & H-twinning            & V-twinning            & H-twinning            & V-twinning            & H-twinning            \\ \hline
MSE                              & 0.136                & 0.135                & 0.137                 & 0.136                 & 0.138                 & 0.135                 & 0.137                 & 0.134                 \\ \hline
MAE                              & 0.257                & 0.255                & 0.257                 & 0.256                 & 0.258                 & 0.256                 & 0.258                 & 0.256                 \\ \hline
NRMSE                            & 0.091                & 0.089                & 0.091                 & 0.089                 & 0.091                 & 0.089                 & 0.090                 & 0.089                 \\ \hline
Initial Mapping (sec.)                             & 642                & N/A               & 1359                 & N/A                & 2628                 & N/A                & 6665                 & N/A                \\ \hline
Update (rounds)                             & N/A               & 3000                & N/A                & 2360                 & N/A                & 2325                 & N/A                & 2302                 \\ \hline
\end{tabular}
\label{table:part}
\vspace{-0.4cm}
\end{table*}

\subsubsection{\textbf{Investigation on Feasibility of Partial Twinning}}

In the context of our VH-Twin mechanism, encompassing both vertical mapping for initialization and horizontal mapping for ongoing updates, our focus is on reducing overall mapping time without compromising twinning accuracy. One potential approach involves decreasing the percentage of twinning network regions covered by BSs in the V-twinning or H-twinning stage. This reduction can be achieved by lowering the participation percentage of BSs during the mapping process, allowing for the exclusion of certain BSs from the construction of C-NDTs. Given the time-series data interdependence among distributed BSs, a randomized selection of a specific percentage of participating BSs in each twinning round is feasible. To validate this approach, Table~\ref{table:part} provides valuable insights into performance variations across multiple dimensions as the participation rate of BSs fluctuates.
Specifically, with participation rates ranging from 40\% to 100\%, we observe the mapping time for V-twinning process increases significantly, from 642 seconds at 40\% to 6,665 seconds at a full twinning level. 
Conversely, in the H-twinning stage, there is a notable decrease in the update overhead as the participation rate increases, attributed to more comprehensive twin initialization. This implies that higher BS participation contributes to more efficient ongoing maintenance of twins, potentially due to reduced redundancy in data synchronization and enhanced data aggregation.
Remarkably, the accuracy of the twin model remains consistently stable across various participation levels in both V-twinning and H-twinning processes. This stability indicates the robustness of our VH-Twin scheme even with fluctuations in the number of participating BSs.
On the other hand, the results reveal the importance of striking a balance between mapping time and update cost. While a more complete level of V-twinning seems advantageous for reducing the following update rounds, it concurrently extends initial mapping time which may consume more resources. This dedicated interplay paves the way for addressing scalability concerns in large-scale network DT systems. Determining an optimal participation rate should involve resource considerations, such as scheduling appropriate bandwidth or slots, to ensure a scalable and efficient twinning process.


\section{Conclusion and Future Work} \label{sec:conclusion}

In this paper, we introduced VH-Twin as a fundamental mapping framework for digitalizing emerging wireless networks.
This framework represents a significant technical advancement, synergistically leveraging the capabilities of vertical twinning and horizontal twinning schemes, along with dynamic connectivity segmentation, to efficient cluster distributed base stations within an urban cellular network.
Extensive evaluations demonstrate that VH-Twin can effectively and efficiently create, deploy, and maintain network DTs.

With the advent of the new digitalization era, such an accurate mapping of network DTs opens up numerous opportunities for future research and development. One avenue for future work involves enhancing the capabilities of DTs to incorporate real-time data streams and predictive analytics, enabling proactive network management and optimization. Additionally, exploring the integration of generative artificial intelligence techniques can further refine DT models, allowing for more precise model creation and simulations. Furthermore, investigating the application of DTs in emerging technologies such as Internet of Things and 6G wireless systems presents promising avenues for integrated intelligence and autonomy.

\section*{Acknowledgment}
This research was supported by the National Science Foundation through Awards CNS--2312138 and CNS--2312139.

\bibliographystyle{IEEEtran}
\bibliography{ref}

\end{document}